\documentclass[aps, prl, reprint, twocolumn,superscriptaddress,floatfix]{revtex4-1}

\usepackage{bbm, amsthm, bm,textcomp, nicefrac,geometry,ragged2e}
\usepackage[dvipsnames]{xcolor}
\usepackage{float}
\usepackage[bbgreekl]{mathbbol}
\usepackage{graphicx,epstopdf,color,verbatim,enumitem,ulem}
\usepackage{svg}
\usepackage{pbox,array}
\usepackage{mathrsfs}
\usepackage{amssymb}
\usepackage{stackrel}
\usepackage{amsmath}
\makeatletter
\usepackage[caption=false]{subfig}
\usepackage[T1]{fontenc}
\usepackage[caption=false]{subfig}
\usepackage{babel}
\addto\captionsenglish{}
\usepackage{dcolumn}    
\usepackage{ifpdf}
\usepackage{lineno,amsfonts}
\usepackage{graphicx}   
\usepackage{bm}         
\usepackage{bbm}
\usepackage{mathrsfs}
\usepackage{upgreek}
\usepackage{mathtools}
\usepackage{epstopdf}
\usepackage{setspace}
\usepackage{float}
\usepackage{natbib}
\usepackage{ifpdf}
\usepackage{qcircuit}
\usepackage{dsfont}
\usepackage{xr}
\usepackage{xr-hyper}
\usepackage{hyperref}

\makeatletter
\newcommand*{\addFileDependency}[1]{
  \typeout{(#1)}
  \@addtofilelist{#1}
  \IfFileExists{#1}{}{\typeout{No file #1.}}
}
\makeatother

\newcommand*{\myexternaldocument}[1]{%
    \externaldocument{#1}%
    \addFileDependency{#1.tex}%
    \addFileDependency{#1.aux}%
}

\myexternaldocument{SM}

\newcommand{\ket}[1]{\vert{#1}\rangle}

\newcommand{\expec}[1]{\langle{#1}\rangle}

\newcommand{\e}[1]{\ensuremath{\times 10^{#1}}}

\newcommand{\luca}{\color{green}}

\begin{document}

\title{A measurement-based variational quantum eigensolver}

\author{R. R. Ferguson}
\thanks{These authors contributed equally.}
\affiliation{Institute for Quantum Computing and Department of Physics and Astronomy, University of Waterloo, Waterloo N2L 3G1, Canada}
\author{L. Dellantonio}
\thanks{These authors contributed equally.}
\affiliation{Institute for Quantum Computing and Department of Physics and Astronomy, University of Waterloo, Waterloo N2L 3G1, Canada}
\author{A. Al Balushi}
\affiliation{Institute for Quantum Computing and Department of Physics and Astronomy, University of Waterloo, Waterloo N2L 3G1, Canada}
\author{K. Jansen} 
\affiliation{NIC, DESY Zeuthen, Platanenallee 6, 15738 Zeuthen, Germany}
\author{W. D\"ur}
\affiliation{Institut f\"ur Theoretische Physik, Universit\"at Innsbruck, Technikerstra{\ss}e 21a, 6020 Innsbruck, Austria}
\author{C. A. Muschik}
\affiliation{Institute for Quantum Computing and Department of Physics and Astronomy, University of Waterloo, Waterloo N2L 3G1, Canada}
\affiliation{Perimeter Institute for Theoretical Physics, Waterloo, Ontario N2L 2Y5, Canada}

\begin{abstract}

Variational quantum eigensolvers (VQEs) combine classical optimization with efficient cost function evaluations on quantum computers. We propose a new approach to VQEs using the principles of measurement-based quantum computation. 
This strategy uses entangled resource states and local measurements. We present two measurement-based VQE schemes. The first introduces a new approach for constructing variational families. The second provides a translation of circuit-based to measurement-based schemes. Both schemes offer problem-specific advantages in terms of the required resources and coherence times.

\end{abstract}

\maketitle

Variational methods are crucial to investigate the physics of strongly correlated quantum systems. Numerical tools like the density matrix renormalization group \cite{perez608197matrix, PhysRevB.88.075133,orus2014advances, PhysRevLett.115.180405} have been applied to several problems including real-time dynamics \cite{PhysRevB.87.115115}, condensed matter physics \cite{PhysRevX.7.031020}, lattice gauge theories \cite{banuls2018tensor, banuls2020review, dalmonte2016lattice, bender2020real}, and quantum chemistry \cite{chan2008introduction,  PhysRevLett.117.210402}. Nevertheless, the classes of states that can be studied with classical computers is limited \cite{PhysRevA.64.022319}. Variational quantum eigensolvers (VQEs) overcome this problem using a closed feedback loop between a classical computer and a quantum processor \cite{mcclean2016theory, farhi2014quantum, preskill2018quantum}. 
The classical algorithm optimizes a cost function -- typically the expectation value of some operator -- which is efficiently supplied by the quantum hardware. The VQE provides an approximation to the (low-lying) eigenvalues of this operator and the corresponding eigenstates. VQEs are advantageous for a variety of applications \cite{shehab2019noise, preskill2018quantum, banuls2020simulating, kaubruegger2019variational, PhysRevLett.120.210501, Jan_article, PhysRevX.6.031045} and have been experimentally demonstrated in fields including chemistry \cite{RevModPhys.92.015003, PhysRevX.6.031007}, particle physics \cite{kokail2019self, paulson2020towards, PhysRevA.100.062319, PhysRevA.100.012320}, and classical optimization \cite{borle2020quantum,bravo2020variational,kyriienko2020solving}.

Existing VQE protocols are based on the circuit model, where gates are applied on an initial state \cite{peruzzo2014variational}. 
These gates involve variational parameters whose optimization allows the resulting output state to approximate the desired target state. We propose a new approach to VQE protocols, based on the measurement-based model of quantum computation (MBQC) \cite{owqc_old_ref, briegel2009measurement, browne2006one, raussendorf_owqc, walther2005experimental,larsen2020deterministic}. In MBQC, an entangled state is prepared and the computation is realized by performing single-qubit measurements.
While the circuit-based and measurement-based models both allow for universal quantum computation and have equivalent scaling of resources \cite{raussendorf_owqc}, they are intrinsically different. The former is limited by the number of available qubits and gates that can be performed, and MBQC by the size of the entangled state one can generate. 
For certain applications, the required coherence times~\cite{article, zwerger2014hybrid} and error thresholds \cite{raussendorf_owqc,article, PhysRevLett.110.260503, zwerger2014hybrid} are much less demanding for MBQC.

%
\begin{figure}
\centering
\includegraphics[width=\columnwidth]{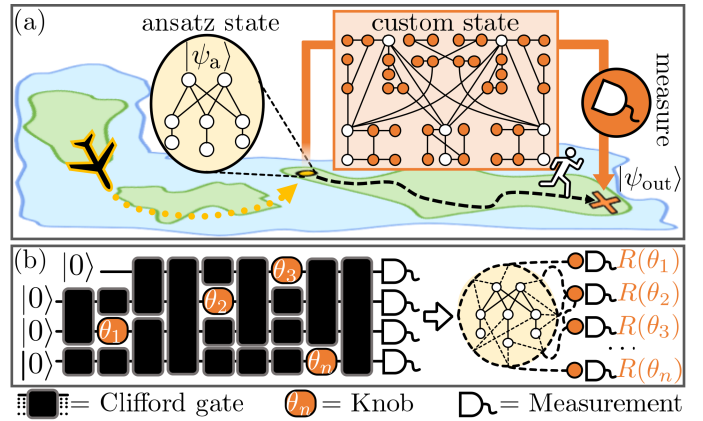}
\caption{\textbf{MB-VQE schemes.} \textit{(a)} Variation of a problem-specific ansatz state by `edge decoration'. An ansatz graph state starts the MB-VQE in a suitable corner of Hilbert space (choice of green island). Next, a classical algorithm explores the neighbourhood (runner on black arrow). The variational optimization exploits a custom state that is obtained by decorating the edges of the ansatz state with auxiliary qubits (orange circles). Their measurement in rotated bases $R(\theta)$ with variational parameters $\theta$ transforms $\lvert \psi_{\rm a}\rangle$ into the output state $\lvert \psi_{\rm out}\rangle$. \textit{(b)} Direct translation of a VQE circuit into a MB-VQE. Left: circuit consisting of Clifford gates (black) and single-qubit parametric gates (`knobs'). Right: corresponding MB-VQE, where the Clifford-part of the circuit has been performed beforehand. The custom state consists of output (white circles) and auxiliary (orange circles) qubits only; the latter are measured in rotated bases and are related to the `knobs' in the circuit.
    }
\label{fig:island_hop}
\end{figure}

Here, we develop a new variational technique based on MBQC, that we call measurement-based VQE (MB-VQE). Our protocols determine the ground state of a target Hamiltonian, which is a prototypical task for VQEs with wide-ranging applications \cite{mcclean2016theory, RevModPhys.92.015003, PhysRevLett.120.210501, PhysRevX.6.031045, moll2018quantum, paulson2020towards, Jan_article}. The underlying idea is to use a tailored entangled state called `custom state', that allows for exploring an appropriate corner of the system's Hilbert space [Fig.~\ref{fig:island_hop}(a)]. This custom state includes auxiliary qubits which, once measured, modify the state of the output qubits. The choices of the measurement bases, and the corresponding variational changes to the state, are controlled by a classical optimization algorithm. This approach differs conceptually and practically from standard VQE schemes since MB-VQE shifts the challenge from performing multi-qubit gates to creating an entangled initial state. 

After presenting the framework for the MB-VQE, we design two specific schemes that are suited to different problem classes. First, we introduce a novel method to construct variational state families, illustrated using the toric code model with local perturbations \cite{wolfgang_tensor_network}. As Fig.~\ref{fig:island_hop}(a) shows, we start from an ansatz state $\lvert \psi_{\rm a} \rangle$ in an appropriate corner of the Hilbert space. To explore this neighbourhood using a classical optimization algorithm, we introduce a custom state and apply measurement-based modifications of $\lvert \psi_{\rm a} \rangle$ that have no direct analogue in the circuit-model. The resulting variational family is not efficiently accessible with known classical methods and is more costly to access with circuit-based VQEs.

Second, we introduce a direct translation of circuit VQEs to MB-VQEs [Fig.~\ref{fig:island_hop}(b)]. Here, the variational state family is the same for the circuit- and the measurement-based approaches, but the implementation differs as the MB-VQE requires different resources and is manipulated by single-qubit measurements only. We exemplify this direct translation for the Schwinger model \cite{schwinger1951theory} and highlight the different hardware requirements and the scaling of resources. As explained below, a translation to MB-VQE is advantageous for circuits containing a large fraction of so-called Clifford gates (e.g. $CX$ gates), as these are absorbed into the custom state.

While MB-VQE is platform agnostic, it opens the door for complex quantum computations in systems where long gate sequences or the realization of entangling gates are challenging. In particular, MB-VQE offers new routes for experiments with photonic quantum systems, thus enlarging the toolbox of variational computations.

\textbf{General framework} The main resource of MBQC are so-called graph states \cite{hein2004multiparty, hein2006entanglement}. 
Graphs as in Fig.~\ref{fig:island_hop} are stabilizer states (eigenstates with $+1$ eigenvalues) of the operators $\hat{S}_n=\hat{X}_n\prod_k\hat{Z}_k$, where $k$ refers to the vertices connected to site $n$.
%
To obtain the desired final state encoded in the output qubits (white circles), single-qubit measurements are performed on auxiliary qubits (orange circles), either in the eigenbasis of the Pauli operators $\hat{X}$, $\hat{Y}$, $\hat{Z}$, or in the rotated basis $R(\theta) \equiv \{(\ket{0} \pm e^{i\theta}\ket{1}) / \sqrt{2}\}$. Depending on the measurement outcomes, the system is probabilistically projected into different states. To make the computation deterministic, so-called byproduct operators and adaptive measurements are required \cite{raussendorf_owqc}. The former applies $\hat{X}$ and $\hat{Z}$ operators to the output qubits depending on the measurement results, while the latter involves adapting the measurement bases $R(\theta)$ based on earlier measurement outcomes. Consequently, adaptive measurements must be performed in a specific order.

An advantage of MBQC is the possibility to simultaneously perform all non-adaptive measurements at the beginning of the calculation (see Fig.~\ref{fig:island_hop}(b) and Supplemental Material (SM) \cite{supp_mat}). This corresponds to the Clifford part of a circuit and includes single- and many-qubit gates. This is independent of the position of the gates in the circuit, and reduces the required overhead and coherence time. Remarkably, this can be either done directly on the graph state in the quantum hardware, or on a classical computer before the experiment. In the latter case, the Gottesmann-Knill theorem \cite{improved_sim} allows for efficiently determining the custom state which is local-Clifford equivalent to the quantum state obtained after all non-adaptive measurements are performed \cite{cliff_on_graph_states}. This state can be directly prepared and used for the MBQC, which may have dramatically fewer auxiliary qubits compared to the initial graph state.

We now explain how MBQC is used to design a MB-VQE. While the classical part of the feedback loop is untouched (the best optimization algorithm \cite{Kingma2014aa, Ruder2017An-overview,Stokes2020Quantum,Audet2006Mesh,Rasmussen2003,Frazier2018aa} is problem dependent \cite{paulson2020towards,cerezo2020variational}), the MB-VQE is based on the creation and partial measurement of a tailored graph state rather than the application of a sequence of gates. Specifically, the quantum part of a MB-VQE comprises an ansatz state $\lvert \psi_{\rm a} \rangle$, a custom state, and a measurement prescription. As schematically represented in Fig.~\ref{fig:island_hop}(a), $\lvert \psi_{\rm a} \rangle$ is a graph state from which we start exploring the variational class of families attainable by the MB-VQE.
%
%
The custom state is then created by expanding $\lvert \psi_{\rm a} \rangle$ into a bigger graph state. This is done by decoration, i.e. by adding new vertices and connecting them to pre-existing sites in the ansatz state. According to a measurement prescription, which is the same at each iteration of the algorithm, the auxiliary qubits of the custom state are then measured, with the remaining ones constituting the output $\lvert \psi_{\rm out} \rangle$ of the quantum processor [see Fig.~\ref{fig:island_hop}(a)]. The cost function to be fed into the classical side of the MB-VQE is then calculated from $\lvert \psi_{\rm out} \rangle$ (e.g. its energy), with the angles $\theta$ of the rotated bases $R(\theta)$ being the variational parameters over which the optimization occurs.


Just like the circuit in a VQE, the custom state determines the success of our MB-VQE. Generally, the more auxiliary qubits are measured in rotated bases $R(\theta)$, the bigger the available class of variational states that can be explored. However, an excessive number of parameters $\theta$ makes the algorithm's convergence slower. Therefore, it is convenient to tailor the custom state to the considered problem. And qubit decoration, with the subsequent measurement of the auxiliary qubits, allows for remarkable control over the desired ansatz state's transformation(s). Not only can one apply gates -- just like in a circuit-based VQE -- by following MBQC prescriptions (see the Schwinger model example), one can also identify completely new patterns of auxiliary qubits, that modify the output state in a way that would be expensive or even impossible with the circuit formalism (see the toric code example). For instance, a single auxiliary qubit measured in $R(\theta)$ and connected to an arbitrary number of output qubits $\lbrace 1,2,3,\dots \rbrace$, acts $e^{i\frac{\theta}{2} \hat{Z}_{1}\otimes \hat{Z}_{2} \otimes \hat{Z}_{3} \otimes \dots}$ onto them \cite{browne2006one}. In a circuit, the same operation requires a linear number of two-qubit gates \footnote{A simple way to get the same operation within the circuit framework is to use an ancilla qubit in place of the auxiliary qubit in the graph, and act with $CX$ gates between the ancilla and the output qubits.}.

\textbf{State variation by edge-modification (perturbed toric code)} Here, we demonstrate how a MB-VQE manipulates states in a different way than a circuit-based VQE. MB-VQEs are advantageous whenever a perturbation $\hat{H}_{\rm p}$ is added to a Hamiltonian $\hat{H}_0$ whose ground state, used as ansatz state $\lvert \psi_{\rm a} \rangle$ below, is a stabilizer state or a graph state.

\begin{figure}
	\centering
	\includegraphics[width=\columnwidth]{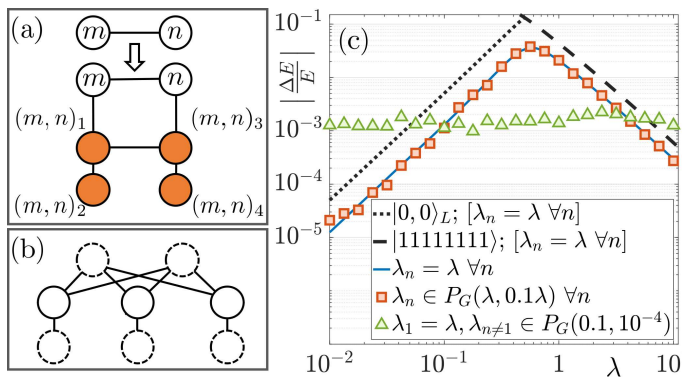}
	\caption{\textbf{Perturbed toric code.} \textit{(a)} Edge modification resource for the MB-VQE [see Fig.~\ref{fig:island_hop}(a)]. Four auxiliary qubits (orange circles), labelled $(m,n)_i$ ($i= 1,\dots,4$), are added to two connected output qubits $m$ and $n$ (white circles). \textit{(b)} Graph state representation of the ansatz state $\lvert 0,0 \rangle_L$. Additional Hadamard gates are applied to qubits with dashed lines. \textit{(c)} Relative difference between the MB-VQE result and the true ground state energy vs the perturbation strength. We let $\lambda_n$ in Eq.~\eqref{eq:perturbation} be equal on all qubits (solid blue line), or sampled from a normal distribution $P_G$ of average $\lambda$ and variance $0.1 \lambda$ (red squares). Green triangles describe a perturbation acting strongly on $\lambda_1$ and weakly on the other qubits. Dotted and dashed lines are computed with respect to $\lvert 0,0 \rangle_L$ (ansatz state) and $\lvert 1 \rangle^{\bigotimes 2N_x N_y}$ (ground state of $\hat{H}_{\rm p}$) for $\lambda_n = \lambda$ $\forall n$.
	}
	\label{fig:toric_code}
\end{figure}

To create the custom state from $\lvert \psi_{\rm a} \rangle$
we employ the pattern in Fig.~\ref{fig:toric_code}(a),
that decorates each connected pair of output qubits $m$ and $n$ with four auxiliary qubits $(m,n)_i$ ($i=1,\dots ,4$), to be measured in rotated bases $R(\theta)$. 
Depending on $R(\theta)$, the entanglement between the qubits $m$ and $n$ is modified, and their state subjected to an additional rotation. For example, if all auxiliary qubits in the custom state are measured with $\theta=0$, we obtain the original ansatz state. However, if all auxiliary qubits are measured with $\theta = \pi/2$, then all entanglement of $\lvert \psi_{\rm a} \rangle$ is eliminated (for more details, see SM \cite{supp_mat}). This decoration technique is tailored to the perturbed toric code example below, in which the ground states of $\hat{H}_{\rm 0}$ and $\hat{H}_{\rm p}$ are maximally entangled and pure, respectively. 
However, it can be easily generalized to expand the class of available variational states (see SM \cite{supp_mat}), thus suiting different scenarios.



%

We apply this MB-VQE approach to the toric code model, 
a quantum error-correcting code defined on a two-dimensional rectangular lattice with periodic boundary conditions \cite{kitaev2003fault}. On the lattice, the number of rows (columns) of independent vertices is $N_x$ ($N_y$) and edges represent qubits. The toric code state is a stabilizer state of so-called star $\hat{A}_s$ and plaquette $\hat{B}_p$ operators. For any vertex $s$ in the lattice, $\hat{A}_s$ acts $\hat{Z}$ on the four incident edges, while $\hat{B}_p$ acts $\hat{X}$ on the four edges in the $p$-th plaquette. The toric code Hamiltonian is then $\hat{H}_{0} = -\sum_s\hat{A}_s -\sum_p\hat{B}_p$. Since $\prod_s\hat{A}_s = \prod_p \hat{B}_p = 1$, the toric code has $2N_xN_y-2$ independent stabilizers, and $\hat{H}_{0}$ has four degenerate ground states $\ket{r,t}_L$ ($r,t = 0,1$), called logical states below. These are simultaneous eigenstates of $\hat{H}_{0}$ and the two logical-$Z$ operators \cite{kitaev2003fault}, as explained in the SM \cite{supp_mat}.

The perturbation added to the toric code Hamiltonian is
\begin{equation}\label{eq:perturbation}
	\hat{H}_{\rm p} = \sum_{n=1}^{2N_x N_y} \lambda_{n} \hat{Z}_n, 
\end{equation}
which corresponds to an inhomogeneous magnetic field. As ansatz state for the MB-VQE we choose the highly entangled graph state $\ket{\psi_{\rm a}} = \ket{0,0}_L$, that approximates the ground state of $\hat{H}_{0} + \hat{H}_{\rm p}$ for small positive values of $\lambda_{n}$. The graph state representation of $\ket{0,0}_L$ can be calculated efficiently classically \cite{cliff_on_graph_states} and is shown  
in Fig.~\ref{fig:toric_code}(b) for $N_x = N_y = 2$.
In SM \cite{supp_mat}, we explain how to adapt our MB-VQE protocol to use an arbitrary superposition of $\ket{r,t}_L$ ($r,t = 0, 1$) as ansatz state, which is more suited for different kinds of perturbations $\hat{H}_{\rm p}$.

Numerical results for the MB-VQE are shown in Fig.~\ref{fig:toric_code}(c). The relative energy difference between the MB-VQE result and the true ground state (calculated via exact diagonalization) is plotted against the perturbation strength. This is done with all $\lambda_n$ in Eq.~\eqref{eq:perturbation} equal to $\lambda$ (solid blue line),  with each $\lambda_n$ drawn from a Gaussian distribution $P_G(\mu,\sigma^2)$ with mean $\mu = \lambda$ and variance $\sigma^2 = 0.1\lambda$ (orange squares), and with $\lambda_1 = \lambda$, $\lambda_n$ randomly sampled from $P_G(\mu = 0.1,\sigma^2 = 10^{-4})$ for $n \neq 1$ (green triangles).
A plot of the infidelity resembles Fig.~\ref{fig:toric_code}(c), with maximum infidelities for the blue curve, red squares and green triangles being 6.2\e{-2}, 6.5\e{-2}, and 9.4\e{-3}, respectively. Fig.~\ref{fig:toric_code}(c) shows that the MB-VQE produces the ground state energy with high confidence when the perturbation strength is very small or very large. Notably, the MB-VQE outperforms the ansatz state (dotted black line) and the ground state of $\hat{H}_{\rm p}$ in Eq.~\eqref{eq:perturbation}
(dashed black line) in all cases. If the perturbation only acts on one qubit, the chosen custom state allows the MB-VQE to find the exact ground state energy within machine precision. This is also the case if the perturbation acts on two disconnected qubits, provided we connect them and add auxiliary qubits as in Fig.~\ref{fig:toric_code}(a). This suggests that the outcome of the MB-VQE can be significantly improved by adding few extra auxiliary qubits.

\textbf{Translating VQEs into MB-VQEs (Schwinger model)} Instead of the approach described above, one can create a MB-VQE by translating the circuit of a VQE into its corresponding custom state and a sequence of measurements. Since a universal set of gates can be realized in a MBQC \cite{raussendorf_owqc}, any VQE can be translated into a MB-VQE. As we discuss below, this strategy is advantageous if the number of parametric adaptive measurements [i.e. ‘knobs’ in Fig. \ref{fig:island_hop}(b)] in the resulting MB-VQE scheme is small.

\begin{figure}
\centering
\includegraphics[width=\columnwidth]{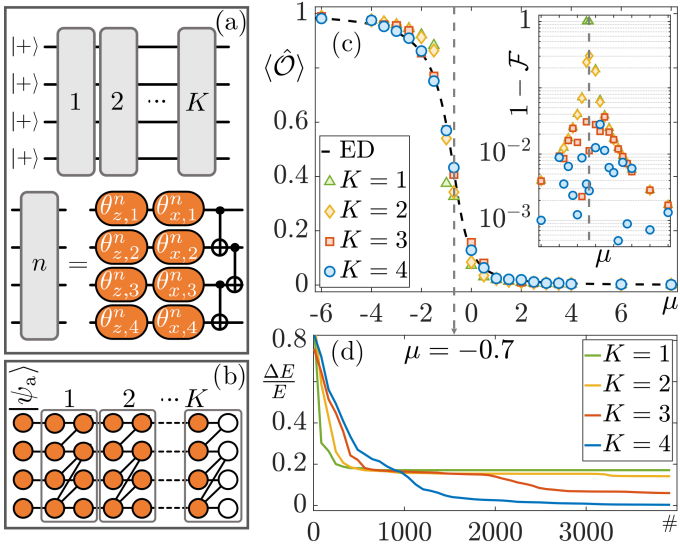}
\caption{\textbf{Schwinger model.} \textit{(a)} Ansatz state and VQE circuit for $S=4$ qubits and $K$ layers. Each layer consists of $CX$ gates and local rotations (orange) parametrized by angles $ \theta_{\nu,i}^{n}$ (with rotation axis $\nu=x,z$; $i=1,\dots,4$). \textit{(b)} MB-VQE custom state for $K$ layers. White circles are output qubits. Auxiliary qubits (orange) are measured in a rotated bases $R(\theta)$. \textit{(c)} The order parameter $\langle{\hat{\mathcal{O}}}\rangle$ vs the fermion mass $\mu$. The dashed line and dots represent exact diagonalization ($ED$) and (MB-)VQE results, respectively, with the number of layers $K$ indicated in the legend. The inset shows the infidelity $1-\mathcal{F}$. \textit{(d)} Relative energy difference $\Delta E / E$ between (MB-)VQE results and $ED$ for $\mu=-0.7$, versus the number 
of iterations in the optimization procedure. The variational parameters are initialized at zero, and $J=\omega = 1$ in Eq.~\eqref{eq:HamSchw}.}
\label{fig:schwinger}
\end{figure}
As an example, we determine the ground state energy of the Schwinger model \cite{schwinger1951theory}, a testbed used for benchmarking quantum simulations in high energy physics \cite{kokail2019self, PhysRevA.98.032331, banuls2020review}. The Schwinger model describes quantum electrodynamics on a one-dimensional lattice and can be cast in the form of a spin model with long range interactions \cite{PhysRevD.56.55, martinez2016real, muschik2017u},
\begin{equation}
\begin{split}
\label{eq:HamSchw}
    \hat{H} & = \frac{J}{2} \sum_{n=1}^{S-2}\sum_{k = n+1}^{S-1} (S-k)\hat{Z}_n\hat{Z}_k - \frac{J}{2} \sum_{n=1}^{S-1}n \text{mod}2 \sum_{k=1}^{n}\hat{Z}_{k} \\
    & + w\sum_{n=1}^{S-1} \left(\hat{\sigma}_n^+ \hat{\sigma}_{n+1}^- + \text{H.C.} \right) + \frac{\mu}{2} \sum_{n=1}^S (-1)^n \hat{Z}_n ,
\end{split}
\end{equation}
where $S$ is the number of fermions, $\mu$ their mass, $w = \frac{1}{2a}$, and $J = \frac{g^2a}{2}$. Here, $a$ and $g$ are the lattice spacing and the coupling strength, respectively, and $\hat{\sigma}^{\pm}_n = (\hat{X}_n \pm i\hat{Y}_n)/2$.

For the VQE protocol, we assume the typical situation where parametric single-qubit gates and fixed entangling gates ($CX$s) are used \cite{arute2020hartree, RevModPhys.92.015003}. We consider a generic VQE circuit, in which a sequence of `layers' is applied \cite{mcclean2016theory}, each containing local rotations and entangling gates. As shown in Fig.~\ref{fig:schwinger}(a) for $S=4$, we choose the layer 
\begin{equation}
    \prod_{n=1}^{S/2-1}CX_{2n, 2n+1} \prod_{n=1}^{S/2}CX_{2n-1, 2n} \prod_{n=1}^S \hat{U}_{x,n}(\theta_{x,n})\hat{U}_{z,n}(\theta_{z,n}),  \label{eq:gates}
\end{equation}
where $\hat{U}_{\nu,n}(\theta_{\nu,n}) = \exp{(i \theta_{\nu,n} \hat{V}_{n}/2)}$ [$(\nu,\hat{V}) = (x,\hat{X})$ or $(\nu,\hat{V}) = (z,\hat{Z})$]. The circuit for the VQE is created by concatenating $K$ layers, where $K$ is big enough to sufficiently explore the relevant subsector of the considered Hilbert space. As described in the SM \cite{supp_mat}, the MB-VQE custom state corresponding to a $K$-layer circuit is obtained by joining the measurement patterns of the gates in Eq.~\eqref{eq:gates}, and performing all non-adaptive measurements classically, which effectively removes the Clifford parts of the circuit. 
The custom state is shown in Fig.~\ref{fig:schwinger}(b). As ansatz state we use $|\psi_{\rm a}\rangle = \bigotimes_{n=1}^{S}\lvert + \rangle$.

The (MB-)VQE simulation results are shown in Fig.~\ref{fig:schwinger}(c) for $S=4$ and different values of $K$. We plot the order parameter $\expec{\hat{\mathcal{O}}} = \frac{1}{2S(S-1)}\sum_{i,j<i}\expec{(1+(-1)^i\hat{Z}_i)(1+(-1)^j\hat{Z}_j)}$ against the fermion mass $\mu$ and correctly observe a second-order phase transition around $\mu=-0.7$ \cite{coleman1976more, kokail2019self, PhysRevA.94.052321}. Increasing $K$ improves the ground state approximation, as demonstrated by the inset in Fig.~\ref{fig:schwinger}(c) and by Fig.~\ref{fig:schwinger}(d). The points near the phase transition require $K\gtrsim 3$ layers ($\gtrsim 28$ qubits), whereas $K=1$ layer ($12$ qubits) suffices for the easiest points. Note that allowing different gates as resources in Eq.~\eqref{eq:gates} generally leads to different convergence rates, as demonstrated by the results in Ref.~\cite{kokail2019self}.

Perfect platforms provided, both the VQE and the MB-VQE give the same result. However, the quantum hardware requirements are different for the two methods. The circuit-based VQE requires $S$ qubits, $2KS$ single-qubit operations, and $K(S-1)$ entangling gates. For the corresponding MB-VQE, a custom state of $S(2K+1)$ qubits and $2KS$ single-qubit operations (measurements) are required. Generally, translating a VQE into its corresponding MB-VQE is advantageous whenever the VQE circuit involves a large Clifford part compared to the number of adaptive measurements (i.e. knobs). In this case, MB-VQE avoids the requirement of performing long gate sequences, which is currently challenging due to error accumulation. This is especially interesting for platforms where entangling gates are hard to realize (e.g. photonic setups) or in systems with limited coherence times.


\textbf{Conclusions} In this paper, we merged the principles of measurement-based quantum computation and quantum-classical optimization to create a MB-VQE. We presented two new types of variational schemes, that are not restricted to our specific examples and can be combined. The first applies when the ansatz state is a stabilizer state. In this case, it is classically efficient to determine the corresponding graph state \cite{cliff_on_graph_states}, which is decorated with additional control qubits and prepared directly. We applied this MB-VQE to the perturbed toric code. Additionally, we showed how to adapt any circuit-based VQE to become a MB-VQE, with the Schwinger model as example.

Experimental proof-of-concept demonstrations can be explored by considering the smallest instance of the planar code \cite{kitaev2003fault} with a perturbation on a single qubit as first step. In this scenario, the MB-VQE requires as few as eight entangled qubits instead of the $44$ used above. Especially promising candidate systems include superconducting qubits, and photonic platforms. The latter recently demonstrated the capability to entangle several thousands of qubits \cite{asavanant2019generation,Larsen369}, and to create tailored graph states \cite{tiurev2020fidelity,tiurev2020high,Schwartz434,PhysRevLett.102.053601}. When designing custom states for future experiments, it will be important to understand the effects of decoherence and it will be interesting to investigate whether MB-VQEs retain the high robustness of MBQC against errors \cite{article, PhysRevLett.110.260503, zwerger2014hybrid}.

Our scheme based on edge decoration provides a new way of thinking about state variations in VQEs. In particular, the effects 
resulting from measuring only one or few entangled auxiliary qubits can be challenging to describe with a simple circuit. The resulting state modifications do not necessarily correspond to unitary operations and can affect a large number of remaining qubits \cite{browne2006one}. Accordingly, MB-VQEs can lead to schemes in which few auxiliary qubits suffice to reach the desired state, while many gates would be required in a circuit-based protocol. Just like circuit optimization in standard VQEs \cite{funcke2020dimensional}, tailored decorations can lead to a leap for MB-VQEs, with the custom state optimized to the specific problem.
The framework presented here provides a starting point for designing VQEs whose properties are different and complementary to the standard approach that is based on varying a state by applying gates.

\textbf{\textit{Acknowledgements}} We thank Jinglei Zhang for proofreading the manuscript, Joachim von Zanthier and Ralf R\"ohlsberger for useful discussions. This work was supported by the Transformative Quantum Technologies Program (CFREF), NSERC and the New Frontiers in Research Fund, the Institute of Quantum Computing (IQC) and the Austrian Science Fund (FWF) through project P30937-N27. We also thank the Army Research Laboratory's Center for Distributed Quantum Information, Cooperative Agreement No. W911NF15-2-0060. CM acknowledges the Alfred P. Sloan foundation for a Sloan Research Fellowship.

%

\end{document}


\title{Supplemental material for: A measurement-based variational quantum eigensolver}

\author{R. R. Ferguson}
\thanks{These authors contributed equally.}
\affiliation{Institute for Quantum Computing and Department of Physics and Astronomy, University of Waterloo, Waterloo N2L 3G1, Canada}
\author{L. Dellantonio}
\thanks{These authors contributed equally.}
\affiliation{Institute for Quantum Computing and Department of Physics and Astronomy, University of Waterloo, Waterloo N2L 3G1, Canada}
\author{A. Al Balushi}
\affiliation{Institute for Quantum Computing and Department of Physics and Astronomy, University of Waterloo, Waterloo N2L 3G1, Canada}
\author{K. Jansen} 
\affiliation{NIC, DESY Zeuthen, Platanenallee 6, 15738 Zeuthen, Germany}
\author{W. D\"ur}
\affiliation{Institut f\"ur Theoretische Physik, Universit\"at Innsbruck, Technikerstra{\ss}e 21a, 6020 Innsbruck, Austria}
\author{C. A. Muschik}
\affiliation{Institute for Quantum Computing and Department of Physics and Astronomy, University of Waterloo, Waterloo N2L 3G1, Canada}
\affiliation{Perimeter Institute for Theoretical Physics, Waterloo, Ontario N2L 2Y5, Canada}

\begin{abstract}

In this supplemental material, we explain the edge decoration used in the main text, we present the notation used for the perturbed toric code example, and review patterns to perform a complete set of quantum gates in MBQC.

\end{abstract}

\maketitle

\section{Edge decoration}\label{sec:edge_modify}

\begin{figure}[t!h!]
\centering
\includegraphics[width = 10 cm]{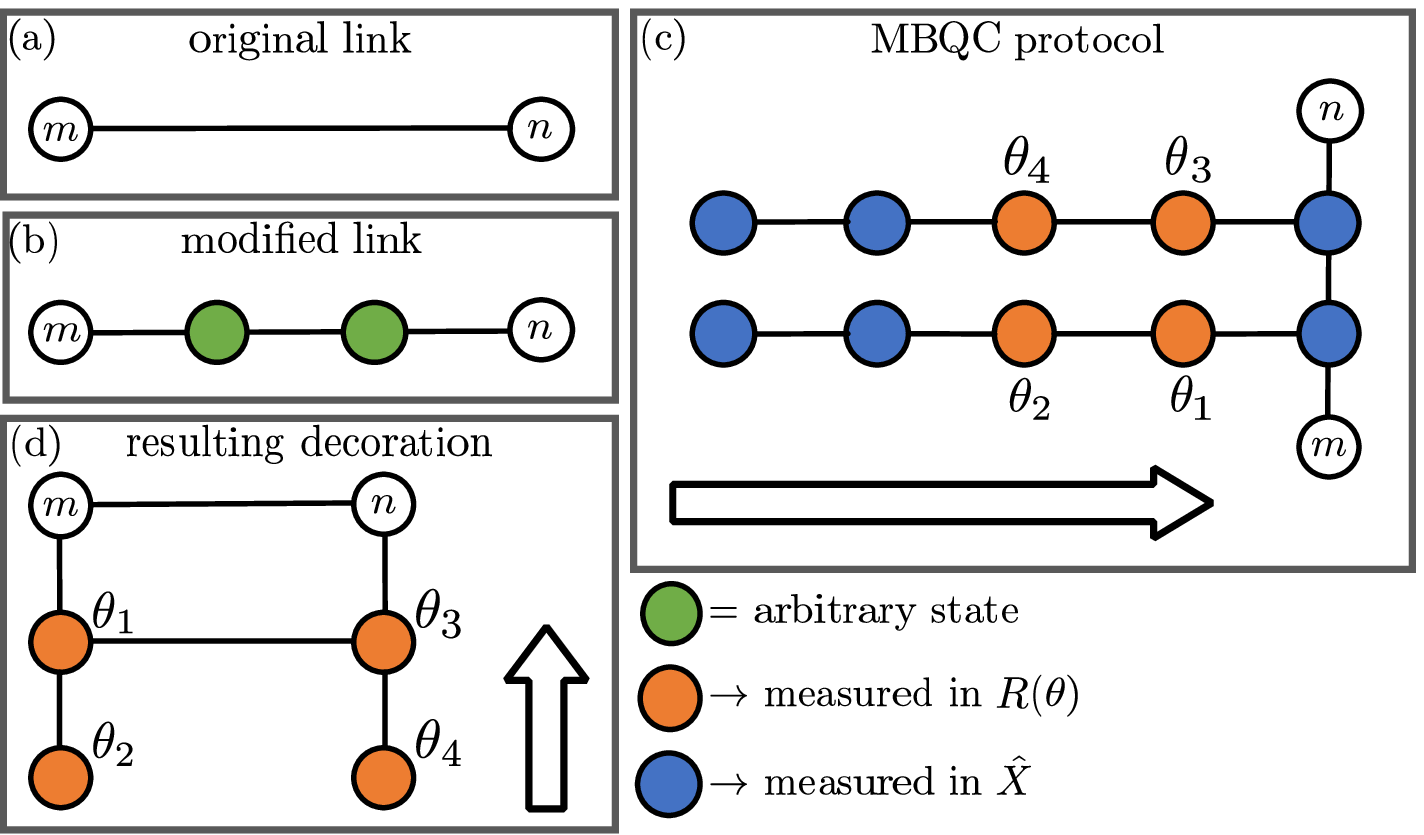}
\caption{\textbf{Edge decoration.} Schematic description of the edge decoration used for the perturbed toric code MB-VQE. The edge connecting a linked pair $(m,n)$ of vertices in $\lvert \psi_{\rm a} \rangle$ [shown in \textit{(a)}] is decorated by adding two auxiliary qubits in an arbitrary state $\lvert \phi \rangle = e^{i\theta_i \hat{X}/2}e^{i\theta_j \hat{Z}/2} \lvert + \rangle$ [shown in \textit{(b)}]. The corresponding MBQC procedure prescribes the pattern in Sec.~\ref{app:measurement_patterns}, resulting in the edge decoration protocol presented in \textit{(c)}. By classically removing Clifford operations, we reduce the number of auxiliary qubits to be added to four, as shown in \textit{(d)} and Fig.~2(a). The arrows represent the temporal order in which the auxiliary qubits have to be measured. In \textit{(c)} and \textit{(d)}, we explicitly indicate the angles $\theta_{i}$ ($i=1,\dots,4$) of the rotated bases in which the corresponding qubits are measured.}
\label{fig:decoration}
\end{figure}

In this section, we explain how we decorate the edges of the perturbed toric code ansatz state $\lvert \psi_{\rm a} \rangle$ [see Fig.~2(a) in the main text] to obtain the custom state. The local perturbation term $\hat{H}_{\rm p}$ in Eq.~(1) leads to a reduction of the amount of entanglement, as can be understood from its ground state, which (for positive weights $\lambda_n$) is the tensor product state $ \bigotimes_n \lvert 1 \rangle_n$. Therefore, we need to decorate the edges of $\lvert \psi_{\rm a} \rangle$ in such a way that, after the measurement of the auxiliary qubits, we both fine-tune the amount of entanglement between connected vertices and rotate the state of the output qubits. This can be done by following the protocol given in Fig.~\ref{fig:decoration}. The idea is to decorate any edge connecting two output qubits $n$ and $m$ with two auxiliary qubits (green circles), which are then measured in the $\hat{X}$ basis. The state in Fig.~\ref{fig:decoration}(b) is prepared by first initializing each of the output qubits in $\lvert+\rangle$ and each of the auxiliary qubit in $\lvert \phi \rangle = e^{i\theta_i \hat{X}/2}e^{i\theta_j \hat{Z}/2} \lvert + \rangle$. Then, for each edge in Fig.~\ref{fig:decoration}(b), we apply a $CZ$ gate to the two connected qubits. The parameters $\theta_i$ and $\theta_j$ describe rotations around the $\hat{X}$ and $\hat{Z}$ axes, respectively, and suffice to prepare each auxiliary qubit in an arbitrary state.

While it is possible (by following the protocol above) to obtain the state depicted in Fig.~\ref{fig:decoration}(b), its wave vector is not stabilized by the operators $\hat{S}_n$ defined in the main text, and as such is not a graph state. Therefore, there are no known methods to obtain a deterministic outcome after the measurement of all auxiliary qubits. A formal MBQC protocol can be obtained by following the patterns presented in Sec.~\ref{app:measurement_patterns} and Ref.~\cite{raussendorf_owqc}. Each of the auxiliary qubits to be used in the decoration is then substituted with five qubits, to be measured in a specific order [see Figs.~\ref{fig:measurement_pattern}(b) and \ref{fig:decoration}(c)]. Since an arbitrary state can be prepared with only two measurements in rotated bases, out of the ten auxiliary qubits in Fig.~\ref{fig:decoration}(c), six can be eliminated \cite{improved_sim}. The resulting state, presented in Fig.~\ref{fig:decoration}(d), only contains four auxiliary qubits per edge of the ansatz state. We remark that the patterns in Figs.~\ref{fig:decoration}(b), \ref{fig:decoration}(c) and \ref{fig:decoration}(d) result in the same output state, provided that the two green auxiliary qubits in Fig.~\ref{fig:decoration}(b) are both projected onto $\lvert + \rangle$ when measured.

We now discuss the effects of the qubit decoration presented in Fig.~\ref{fig:decoration}. When the auxiliary qubits are prepared in the state $\lvert \phi \rangle = \lvert + \rangle$, their measurement results in the identity transformation, and we recover the original state as in Fig.~\ref{fig:decoration}(b). However, if both auxiliary qubits are either in $\lvert \phi \rangle =\lvert 0 \rangle$ or $\lvert \phi \rangle =\lvert 1 \rangle$, their measurement eliminates all pre-existing entanglement between the original output qubits and we obtain a separable state. To better understand the effects of our decoration, we look at the state $\lvert \psi_{n,m} \rangle$ of the two output qubits $n$ and $m$ after the auxiliary ones are measured. With the parameters $\theta_i$ ($i=1,\dots,4$) introduced in Fig.~\ref{fig:decoration}, we find
\begin{equation}\label{eq:stateafterdec}
\begin{split}
   \lvert \psi_{n,m} \rangle = &  \left[ 1 + \cos\theta_4\sin\theta_1\sin\theta_2 + \cos\theta_1\cos\theta_3\sin\theta_2\sin\theta_4 + \cos\theta_2\sin\theta_3\sin\theta_4 \right] \lvert 0 \rangle_m \lvert 0 \rangle_n \\
    & + \left[ \cos^2\frac{\theta_4}{2} + \frac{1}{2} \left( -1 + \cos\theta_4 \right) + \sin\theta_1\sin\theta_2 + i \sin\theta_4 \left( \cos\theta_2\cos\theta_3 - \cos\theta_1 \sin\theta_2 \sin\theta_3 \right) \right] \lvert 0 \rangle_m \lvert 1 \rangle_n \\
    & + \left[ \cos\theta_2 +\sin\theta_3\sin\theta_4 + i \sin\theta_2 \left( \cos\theta_1\cos\theta_4 - \cos\theta_3\sin\theta_1\sin\theta_4 \right) \right]  \lvert 1 \rangle_m \lvert 0 \rangle_n \\
    & + \left[ -\cos\theta_2\cos\theta_4-i\cos\theta_1\sin\theta_2 + \sin\theta_4 \left( -i\cos\theta_3 + \sin\theta_1\sin\theta_2\sin\theta_3 \right) \right] \lvert 1 \rangle_m \lvert 1 \rangle_n ,
\end{split}
\end{equation}
which, for simplicity, is not normalized.
From this equation, it is possible to see that for $\theta_i = 0$ ($\theta_i = \pi/2$) $\forall$ $i$, we get the initial graph state $\lvert \psi_{n,m} \rangle = CZ \lvert + \rangle_m \lvert + \rangle_n$ (the pure state $\lvert \psi_{n,m} \rangle = \lvert + \rangle_m \lvert + \rangle_n$).
For arbitrary angles $\theta_i$, the entanglement between the output qubits is generally reduced, and their wave vector rotated according to Eq.~\eqref{eq:stateafterdec}. By taking a a deeper look on Eq.~\eqref{eq:stateafterdec}, it is possible to realize that the decoration technique presented here does not allow for reaching an arbitrary output state $\lvert \psi_{n,m} \rangle$. For instance, there is no combination of the angles $\lbrace\theta_1,\dots,\theta_4\rbrace$ such that $\lvert \psi_{n,m} \rangle = \lvert 0 \rangle_m \lvert 1 \rangle_n$. This limitation can be lessened or even removed in several ways. For instance, adding a third qubit in an arbitrary state on the link connecting $m$ and $n$, allows for more control over the output state $\lvert \psi_{n,m} \rangle$. Otherwise, after the auxiliary qubits are measured, one can rotate the states of the two output qubits by acting single qubit rotations onto them, as explained in Sec.~\ref{app:measurement_patterns}.

We remark that the number of auxiliary qubits can be further reduced by switching from a deterministic to a probabilistic protocol. While the decoration presented in Fig.~\ref{fig:decoration}(b) does not resort to a graph state, it can still be used for a MB-VQE. From one side, this saves half the auxiliary qubits that are needed for the creation of the custom state. From the other side, the loss of determinism in the outcome forces to repeat each iteration of the MB-VQE algorithm until a specific measurement outcome is obtained. The number of extra repetitions that one must perform scales exponentially with the number of auxiliary qubits, making this probabilistic approach useful for toy examples only.


\section{Toric Code and Logical States}\label{sec:jump_op}

\begin{figure}[t!h!]
\centering
\includegraphics[width = 10cm]{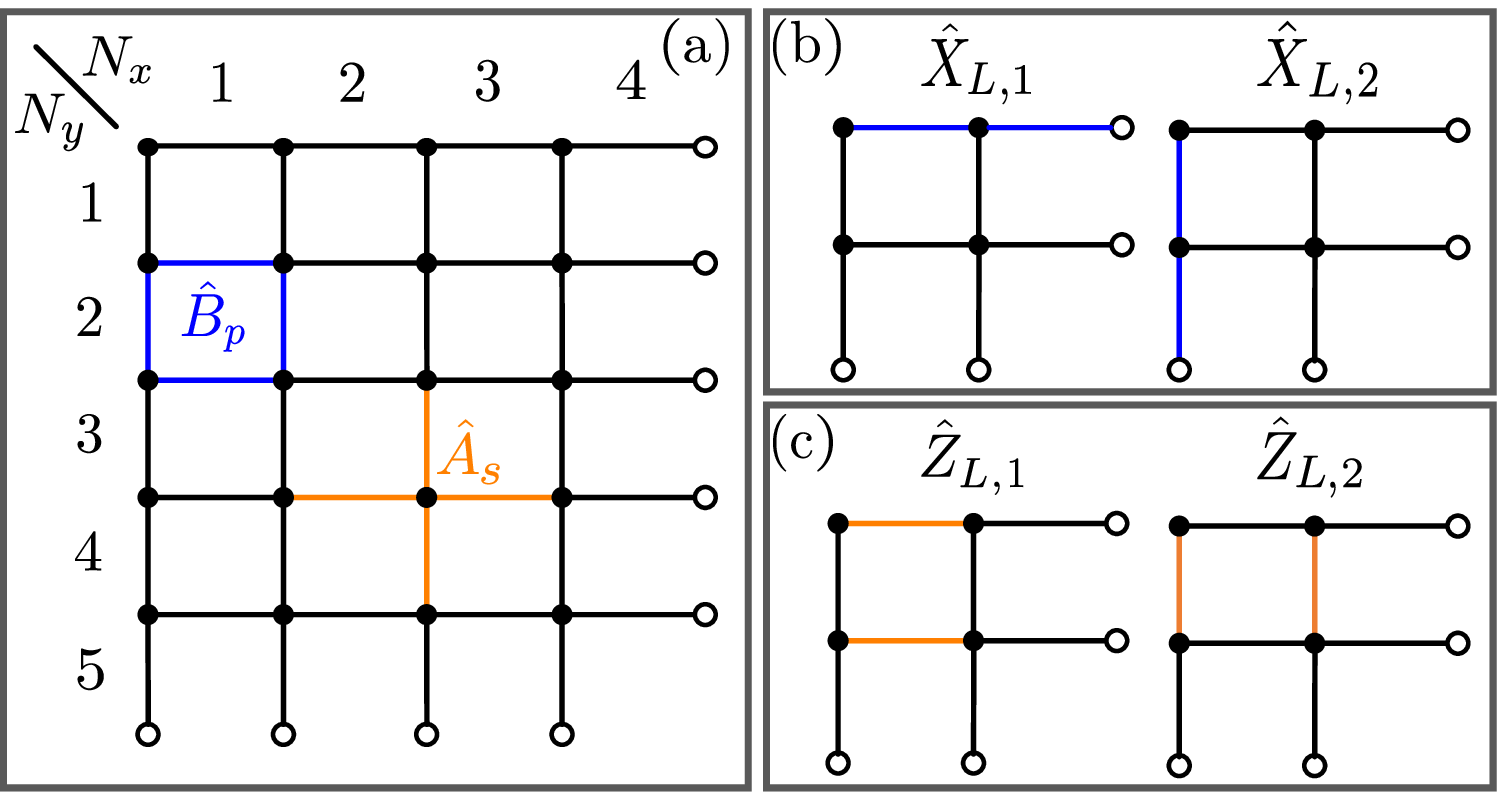}
\caption{\textbf{Toric code notation.} \textit{(a)} Toric code for $N_x = 4$ and $N_y = 5$. Qubits exist on the edges of the lattice, and two of the generators $\hat{A}_s$ and $\hat{B}_p$ are explicitly represented. Since the lattice lies on a torus, periodic boundary conditions are enforced, as shown by empty dots. Schematic representations of the logical-$X$ and logical-$Z$ operators in the case $N_x=N_y=2$ are given in \textit{(b)} and \textit{(c)}, respectively. In the whole figure, links colored in blue (orange) represent the action of the $\hat{X}$ ($\hat{Z}$) operator on the corresponding qubit.}
\label{fig:toric_code_explained}
\end{figure}

In this section, we describe in more detail the notation for the toric code, and explain how to prepare a general MB-VQE ansatz state consisting of an arbitrary superposition of the four degenerate states $\lvert r,t \rangle_{L}$ ($r,t=0,1$). Referring to Fig.~\ref{fig:toric_code_explained}(a), the toric code is defined on a two-dimensional lattice with periodic boundary conditions. Qubits lie on the edges, and the ground state is stabilized by the plaquette $\hat{B}_{p}$ and star $\hat{A}_s$ operators. Since there are two more degrees of freedom than independent stabilizers, the toric code is characterized by four so-called logical states $\lvert r,t \rangle_{L}$ ($r,t=0,1$), corresponding to the degenerate ground states of the toric code Hamiltonian. To explicitly write down these logical states, we add two extra stabilizers to the set of operators  $\hat{A}_s$ and $\hat{B}_p$. There are two common choices for the pair of extra stabilizers to be added, which are called logical-$X$ ($\hat{X}_{L,1}$ and  $\hat{X}_{L,2}$) and logical-$Z$ ($\hat{Z}_{L,1}$ and  $\hat{Z}_{L,2}$), and are shown in Figs.~\ref{fig:toric_code_explained}(b) and \ref{fig:toric_code_explained}(c), respectively. Explicitly, $\hat{Z}_{L,1}$ ($\hat{Z}_{L,2}$) acts $\hat{Z}$ on all horizontal (vertical) edges of an arbitrarily chosen column (row). Similarly, $\hat{X}_{L,1}$ ($\hat{X}_{L,2}$) acts $\hat{X}$ on all horizontal (vertical) edges of an arbitrarily chosen row (column). Both logical-$X$ and logical-$Z$ operators commute with all the other plaquette $\hat{B}_p$ and star $\hat{A}_s$ operators, but do not generally commute between themselves.
%


The logical state $\lvert r,t \rangle_L$ can then be defined as the unique ground state of $\hat{H}_{0} - (-1)^r\hat{Z}_{L,1} - (-1)^t\hat{Z}_{L,2}$ ($r,t = 0,1$). As mentioned in the main text and shown in Fig.~\ref{fig:toric_code_explained}(c), in the considered situation where the weights of the perturbation $\hat{H}_{\rm p}$ in Eq.~(1) are small and positive, $\lvert 0,0 \rangle_L$ approximates well the ground state of the perturbed Hamiltonian $\hat{H}_0 + \hat{H}_{\rm p}$ (this can be proven with perturbation theory). As such, $\lvert 0,0 \rangle_L$ can be used as ansatz state in a successful MB-VQE. However, if some of the weights $\lambda_n$ are negative and/or the perturbation $\hat{H}_{\rm p}$ is changed to include other operators -- like $\hat{X}$ -- then $\lvert 0,0 \rangle_L$ is not a good ansatz state anymore. Generally, the best ansatz state becomes a superposition $\sum_{r,t} c_{r,t} \lvert r,t \rangle_L$, with $\sum_{r,t} \lvert c_{r,t} \rvert^2 = 1$ ($r,t=0,1$). To create this superposition, it is possible to use the generic $SU(4)$ rotation described in Ref.~\cite{su4_article}, with the logical-$X$ and logical-$Z$ operators as generators. Importantly, this $SU(4)$ rotation can be implemented in a MBQC fashion using the tools described in Ref.~\cite{browne2006one}. The seven parameters describing this $SU(4)$ rotation can then be used as input parameters in the MB-VQE, such that the algorithm itself becomes capable of finding the best ansatz state within the degenerate subsector containing the four logical states.


\section{Examples of Measurement Patterns\label{app:measurement_patterns}}

In this section, we summarize the results of Ref.~\cite{raussendorf_owqc} which are relevant to the proposals put forward in the main text. In particular, we review how a universal set of gates can be realized in the MBQC framework. 

\begin{figure}
\centering
\includegraphics[width = 10cm]{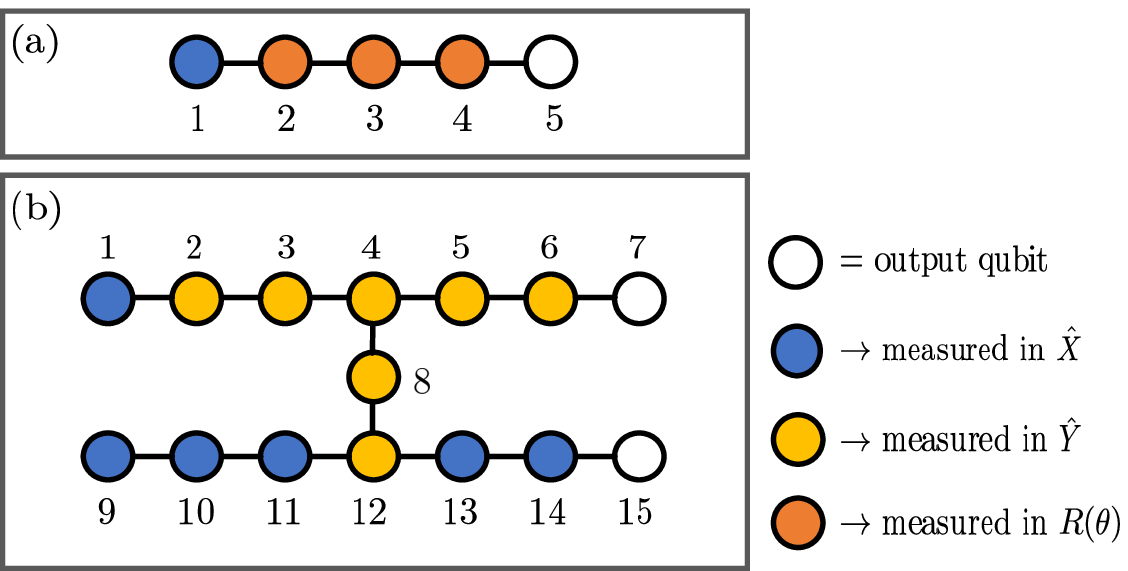}
\caption{\textbf{Set of gates in MBQC.} \textit{(a)} Measurement pattern for a general single-qubit unitary operation. Qubit $1$ is the input qubit. \textit{(b)} Measurement pattern for the $CX$ gate. Qubits $1$ and $9$ are the input qubits.}
\label{fig:measurement_pattern}
\end{figure}

In the following, we denote the measurement result on qubit $n$ as $(-1)^{s_n}$, where $s_n \in \{0,1\}$. A generic single-qubit unitary gate can be written as
\begin{equation}\label{eq:unitary}
    \hat{U}(\theta_1,\theta_2, \theta_3) = \hat{U}_x(\theta_3) \hat{U}_z(\theta_2)\hat{U}_x(\theta_1),
\end{equation}
which corresponds to a rotation of an arbitrary angle around an arbitrary axis.
The measurement pattern realizing this gate is shown in Fig.~\ref{fig:measurement_pattern}(a), where qubit $1$ is the input qubit in a state $\lvert {\rm a} \rangle$, qubits $2$, $3$, and $4$ are measured in the bases $R(-(-1)^{s_1}\theta_1)$, $R(-(-1)^{s_2}\theta_2)$, and $R(-(-1)^{s_1+s_3}\theta_3)$, respectively, and qubit $5$ is the output qubit. Measuring the qubits in this order results in the final state of the output qubits $\hat{U}_{\Sigma}\hat{U}(\theta_1, \theta_2, \theta_3)\lvert {\rm a} \rangle$, where the byproduct operator $\hat{U}_{\Sigma}$ is
\begin{equation}
    \hat{U}_{\Sigma} = \hat{X}^{s_2+s_4}\hat{Z}^{s_1+s_3}.
\end{equation}
This is an example of adaptive measurements, since the measurement bases of qubits $2$, $3$, and $4$ depend on previous results. By contrast, a $CX$ gate acting on two qubits [see Fig.~\ref{fig:measurement_pattern}(b)] corresponds to a non-adaptive measurement pattern. In this case, the measurements can be performed in any order or simultaneously. The byproduct operators for the $CX$ gate are
\begin{equation}
\begin{split}
    \hat{U}_\Sigma = &\hat{X}_7^{s_2+s_3+s_5+s_6} \hat{Z}_7^{s_1+s_3+s_4+s_5+s_8+s_9+s_{11}+1}  \\ &\hat{X}_{15}^{s_2+s_3+s_8+s_{10}+s_{12}+s_{14}} \hat{Z}_{15}^{s_9+s_{11}+s_{13}},
    \end{split}
\end{equation}
and the output qubits are qubits $7$ and $15$.

While Fig.~\ref{fig:measurement_pattern} explains how to implement individual gates in the MBQC formalism, one can implement a sequence of gates by concatenating the corresponding measurement patterns. In the remainder of this section, we explain how to merge measurement patterns. Assume that $\hat{A}$ and $\hat{B}$ are two gates, corresponding to two graph states with input, output and auxiliary qubits. In order to perform $\hat{B}\hat{A}$ in the MBQC framework, we combine the measurement patterns of $\hat{A}$ and $\hat{B}$ such that the input qubits of $\hat{B}$ are the same as the output qubits of $\hat{A}$. As a result, $\hat{B}\hat{A}$ has the same input qubits as $\hat{A}$ and the same output qubits as $\hat{B}$. Let $\hat{U}_{\sigma A}$ and $\hat{U}_{\sigma B}$ be the respective byproduct operators and assume that $\hat{B}$ is a Clifford gate. Then, the overall byproduct operator is 
$\hat{U}_\Sigma = \hat{U}_{\sigma B} \hat{U}'_{\sigma A}$, where
\begin{subequations}
	\begin{align}
	&\hat{U}_{\sigma B}\hat{B}\hat{U}_{\sigma A}\hat{A} =  \hat{U}_{\sigma B}\hat{U}_{\sigma A}'\hat{B}\hat{A}, \\
    &\hat{U}_{\sigma A}' =   \hat{B}\hat{U}_{\sigma A}\hat{B}^\dagger.
	\end{align}
\end{subequations}
If instead $\hat{B}$ is a rotation gate [as in Eq. \ref{eq:unitary}],  the overall byproduct operator is $\hat{U}_\Sigma = \hat{U}_{\sigma B} \hat{U}_{\sigma A}$, and
\begin{subequations}
	\begin{align}
	&\hat{U}_{\sigma B}\hat{B}\hat{U}_{\sigma A}\hat{A} = \hat{U}_{\sigma B}\hat{U}_{\sigma A}\hat{B}'\hat{A}, \\
    &\hat{B}' =  \hat{U}_{\sigma A}^{\dagger}\hat{B}\hat{U}_{\sigma A}.
	\end{align}
\end{subequations}
This modification of $\hat{B}$ affects the bases of the rotated measurements, which have to be fixed following the protocol in Ref.~\cite{raussendorf_owqc}. 
%
